\documentclass[12pt,a4paper]{article}

\usepackage{a4wide}

\usepackage{latexsym}
\usepackage{epsf}
\usepackage{amssymb}
\usepackage{graphicx}
\usepackage{amsmath, cite}
\usepackage{amsmath,amssymb,amsthm}
\usepackage{verbatim}
\usepackage{hyperref}
\usepackage{amsmath}
\usepackage{amsfonts}
\usepackage{xfrac}
\usepackage{slashed}
\usepackage{tikz}
\usepackage{cancel}
\usepackage{bbm}
\usepackage{cite}

\usepackage{fancyhdr}
\usepackage{datetime}

\renewcommand{\d}{\textrm{d}}

\newcommand{\w}{\wedge}

\newcommand\varpm{\mathbin{\vcenter{\hbox{%
  \oalign{\hfil$\scriptstyle+$\hfil\cr
          \noalign{\kern-.3ex}
          $\scriptscriptstyle({-})$\cr}%
}}}}

\newcommand\varmp{\mathbin{\vcenter{\hbox{%
   \oalign{\hfil$\scriptstyle-$\hfil\cr
           \noalign{\kern-.3ex}
          $\scriptscriptstyle({+})$\cr}%
}}}}

\newcommand{\be}{\begin{equation}}
\newcommand{\ee}{\end{equation}}
\newcommand{\bea}{\begin{eqnarray}}
\newcommand{\eea}{\end{eqnarray}}
\newcommand{\C}{\mathbb{C}}

\usetikzlibrary{arrows,automata,positioning,calc,trees,decorations.pathmorphing,decorations.markings}

\usepackage[latin1]{inputenc}
\usepackage{aurical}
\usepackage{anyfontsize}

\begin{document}
\numberwithin{equation}{section}
\begin{flushright}
\small
IPhT-T16/071
\normalsize
\end{flushright}

\vspace{0.5 cm}

\begin{center}
{\fontsize{45}{40}\selectfont \textbf{\Fontamici There and back again:\\ \vspace{0.1cm} A T-brane's tale}}

\vspace{2.6 cm} {\large Iosif Bena, Johan Bl{\aa}b{\"a}ck, Ruben Minasian, Raffaele Savelli}\\

\vspace{0.85 cm}{Institut de Physique Th{\'e}orique, Universit{\'e} Paris Saclay, CEA, CNRS,\\ Orme des Merisiers, F-91191 Gif-sur-Yvette, France}
\\\vspace{0.4 cm}

\vspace{0.45cm} {\small\upshape\ttfamily iosif.bena, johan.blaback, ruben.minasian, raffaele.savelli @ cea.fr} \\

\vspace{3cm}


\textbf{Abstract}
\end{center}

\begin{quotation}

T-branes are supersymmetric configurations described by multiple D$p$-branes with worldvolume flux and non-commuting vacuum expectation values for two of the worldvolume scalars. When these values are much larger than the string scale this description breaks down. We show that in this regime the correct description of T-branes is in terms of a single D$p$-brane, whose worldvolume curvature encodes the T-brane data. We present the tale of the journey to reach this picture, which takes us through T-dualities and rugby-ball-shaped brane configurations that no eye has gazed upon before.

\end{quotation}

\thispagestyle{empty}
\newpage


\tableofcontents


\section{Introduction}\label{sec:intro}

T-branes were first introduced \cite{Cecotti:2010bp} as special types of D7-brane vacuum configurations, where the eigenvalues of the complex worldvolume scalar $\Phi$ fail to capture the physics of the system (see also the earlier work \cite{Donagi:2003hh}). As a consequence, the profile of $\Phi$ is always entangled with a non-trivial worldvolume flux, and the spectrum of low-energy fluctuations typically features an interesting variety of unconventional brane-model-building phenomena. T-branes are not really special to D7-branes, but also exist for other D$p$-branes (as follows trivially from T-duality).

Since their introduction, there has been a stream of efforts on uncovering the peculiarities of these supersymmetric vacua from multiple perspectives, and on investigating their potential applications to string phenomenology \cite{Chiou:2011js,Donagi:2011jy,Donagi:2011dv,Marsano:2012bf,Anderson:2013rka,DelZotto:2014hpa,Collinucci:2014qfa,Collinucci:2014taa,Cicoli:2015ylx,Carta:2015eoh,Collinucci:2016hpz}. There is however an aspect of T-branes that so far has not been thoroughly investigated: The key feature of T-branes is the presence of a non-trivial worldvolume scalar commutator $[\Phi,\Phi^\dagger]$, hinting to possible connections \cite{Cecotti:2010bp, DelZotto:2014hpa} to other D-brane physics where non-Abelian effects become important, such as the dielectric (Myers) effect \cite{Myers:1999ps} or the non-Abelian realization \cite{Constable:1999ac} of the Callan-Maldacena monopole \cite{Callan:1997kz}.

However, it is unclear what this connection is. Non-Abelian fields on D$p$-branes normally give rise to dielectric D$(p+2)$-brane charges \cite{Myers:1999ps, Constable:1999ac}, and this happens when {\em three} of their real worldvolume scalars have non-commuting expectation values. In contrast, for T-branes only two scalars are non-commuting, which makes the connection tenuous. Furthermore, for D7-branes the only possible dielectric dipole charge corresponds to D9-branes. Since these branes are space-filling, they have infinite energy density from a D7 worldvolume perspective and therefore cannot be thought of as dielectric branes.

There is another way to phrase this problem. The Myers effect relates a non-Abelian D$p$-brane description of a system to a description in terms of an Abelian D$(p+2)$-brane that wraps a two-sphere and has worldvolume flux.  The regimes of validity of the two descriptions are complementary. The non-Abelian description is valid when the commutators of the expectation values of the fields are small compared to the string scale, while the Abelian D$(p+2)$-brane description is valid when these commutators are large (and hence the curvature of the two-sphere wrapped by the D$(p+2)$-brane is lower than the string scale) \cite{Myers:1999ps, Constable:1999ac}. As we explained above, for T-branes the naive Abelian description in terms of dielectric branes is problematic. Therefore, we have no way to describe them when their worldvolume fields are large and the non-Abelian picture breaks down. The goal of this paper is to construct such a description.

The final result is that our Abelian description of the T-brane does not involve D$p$-branes polarizing into D$(p+2)$-branes, but rather a single curved D$p$-brane, whose worldvolume curvature encodes the original flux (or equivalently the non-commutative) data of the T-brane. Focusing (without loss of generality) on D7-branes, we explicitly show that the easiest possible T-brane configuration preserving minimal supersymmetry in $5+1$ dimensions corresponds to a single D7-brane with vanishing worldvolume flux, extended along a specific holomorphic curve in the four remaining directions.\footnote{Our result is reminiscent of brane recombination for T-branes \cite{Cecotti:2010bp,Collinucci:2014qfa}, but there is a crucial difference: The effect we find is a large-$N$ effect, which cannot be captured by a gauge that only retains the holmorphic data (see Section \ref{sec:conc} for further remarks).}

To reach our result we make a detour through a T-dual type IIA system consisting of D6-branes ending on a D8-brane, for which the relation between the non-Abelian \cite{Constable:1999ac} and the Abelian  \cite{Callan:1997kz} description is well understood. Our T-brane tale is schematically depicted in Figure \ref{MasterFigure}: We start with the traditional description of T-branes via Hitchin-like equations for a stack of D7-branes, which relate the non-commuting worldvolume scalars to a non-trivial worldvolume flux. We T-dualize this configuration along one of the D7 worldvolume directions aligned with the worldvolume flux; this transforms the T-brane into a smeared stack of D6-branes satisfying Nahm-like equations.\footnote{The duality between Hitchin's and Nahm's systems was recently exploited in \cite{DelZotto:2014hpa} to investigate the relationship between T-branes and D8/D6 systems with a Nahm pole.} We then construct the Abelian description of this system, in terms of a funnel-shaped D8-brane with a rugby-ball cross-section and non-trivial worldvolume flux (which can be thought of as a stretched Callan-Maldacena spike \cite{Callan:1997kz}). This description becomes more and more accurate as the number of D6-branes increases. Our final step is to return to type IIB, by T-dualizing this Abelian D8-brane back along the same direction. This results in a single D7-brane extended along a two-dimensional holomorphic manifold, which we claim to be the correct description of T-branes away from the non-Abelian regime. As an immediate consistency check of our result, the fact that T-branes and D7-branes have compatible supersymmetries is encoded in the holomorphicity of this two-dimensional manifold.

\begin{figure}[h!]
  \begin{center}
  \includegraphics{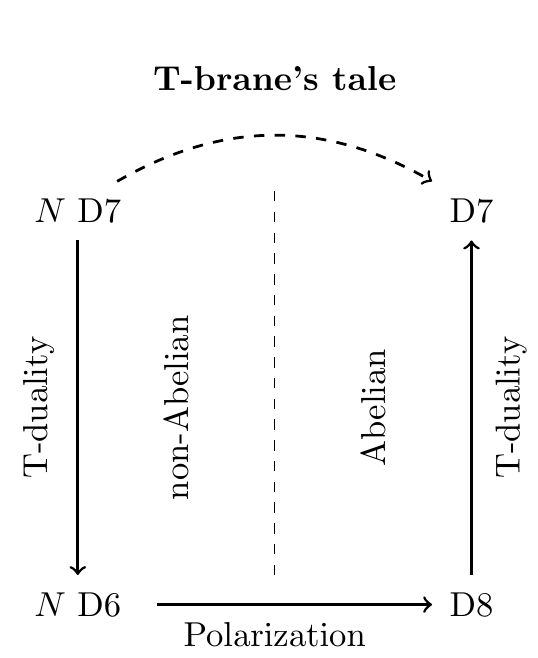}
  \end{center}
  \caption{Summary of our construction.}
\end{figure}\label{MasterFigure}

Along the way we find a novel, two-parameter family of D8-brane solutions with funnel shape, whose cross-section is a generic tri-axial ellipsoid. For a particular limit of the parameters, this family of vacua degenerates to the well-known Callan-Maldacena solution with spherical cross-section \cite{Callan:1997kz}. This class of solutions is interesting in its own right,  and deserves further study. For example, it is likely that such general solutions will encode more general T-brane configurations than the one we considered here, which corresponds to a one-parameter bi-axial ellipsoid (rugby ball) sub-class.

The organization of the paper follows the development of the tale. We start by briefly reviewing in Section \ref{sec:T} the traditional presentation of T-branes, specifying the particular context and vacuum configurations we will be focusing on. In Section \ref{sec:D6} we discuss the first T-duality and describe the vacuum solution in the non-Abelian language of D6-branes. In Section \ref{sec:D8} we describe the same physical system using the Abelian Born-Infeld action of a funnel-shaped D8-brane with rugby-ball cross-section. Section \ref{sec:back} is the epilogue of the tale, where we T-dualize the system back to the original duality frame and obtain the description of the T-brane in terms of a curved D7-brane. This is the main result of this paper. In Section \ref{sec:conc} we make some concluding remarks and discuss our hopes for future adventures. In Appendix \ref{app:D8} we construct the D8-brane solutions with generic tri-axial ellipsoidal cross-section.

\section{T-branes}\label{sec:T}

Our T-brane tale starts in this section, where we introduce its main character: A specific supersymmetric vacuum configuration of type IIB string theory preserving a quarter of the original supercharges and six-dimensional Poincar\'e invariance. We refer to \cite{Beasley:2008dc,Cecotti:2010bp,Anderson:2013rka} for background material relevant to the present discussion.

Consider type IIB string theory on $\mathbb{R}^{5,1} \times \C^2$, where $\C^2$ is parametrized by coordinates $w,z$, and a stack of $N$ D7-branes placed at $z=0$.\footnote{The $5+1$ space-time dimensions play no role in our tale, and will hence not be discussed.} The BPS conditions for the D7-branes are given by a set of complex equations and a set of real ones, given respectively by
\begin{subequations}\label{HitchinSystem}
\bea
\bar{\partial}_A\,\Phi&=&0\,, \label{HF}\\
F+[\Phi,\Phi^\dagger]&=&0\,.\label{HD}
\eea
\end{subequations}
Here $\Phi$ is the Higgs field made from the two worldvolume scalars of the D7, which transforms in the adjoint of the gauge group $G=SU(N)$.\footnote{The field $\Phi$ should be regarded as a $(1,0)$-form on the $w$-plane, as reflected in equations \eqref{HitchinSystem}, but in our non-compact setting, this is not essential. We also disregard the $U(1)$ associated to the center of mass, which decouples.} When $\Phi$ is diagonal, its eigenvalues give the transverse positions (in the $z$-plane) of the components of the D7 stack.
The anti-holomorphic covariant derivative in the $w$ coordinate is $\bar{\partial}_A \equiv  \bar{\partial}+[A_{\bar{w}},\cdot]$, and $A_{\bar{w}}=-(A_w)^\dagger$ is the anti-holomorphic part of an anti-Hermitian $SU(N)$ gauge connection with (self-adjoint) field strength $F=\partial A_{\bar{w}}-\bar{\partial}A_w+[A_w,A_{\bar{w}}]$. Equations \eqref{HitchinSystem} are known as the Hitchin system \cite{Hitchin:1986vp}.\footnote{
It is possible to couple this system to defects of the worldvolume effective theory, which show up as a triple of adjoint-valued moment maps in the right hand side of \eqref{HitchinSystem}. The latter can be physically interpreted as vacuum expectation values of bifundamental matter fields localized on the points of $\C^2$ where the D7 stack intersects other 7-branes \cite{Anderson:2013rka}. In this paper we will just consider an isolated stack of D7-branes, and thus not allow for this possibility.}

The vacuum solution we want to examine here is specified in terms of $N-1$ real functions on the complex $w$-plane: $\{f_a\}_{a=1,\ldots,N-1}$. The Higgs field is a nilpotent $N\times  N$ matrix of the form\footnote{We adopt the convention of summing over repeated indices and we work in units of $2\pi\alpha'$.}
\be\label{Phiu}
\Phi=\left(\begin{array}{cccccc} 0&\phi_1& & &\\ &0 & \phi_2 & & 0 \\ & & \ddots&\ddots& \\ &0&&0&\phi_{N-1}\\&&&&0 \end{array}\right)\,,\qquad \phi_a=\sqrt{a(N-a)}\,e^{C_{ab}f_b/2}\,,
\ee
with $C_{ab}$ the Cartan matrix of $SU(N)$. The gauge field $A$ is diagonal, with anti-holomorphic part given by
\be\label{GaugeField}
A_{\bar{w}}=-\frac{1}{2}\bar{\partial}f_a\,C_a\,,\qquad (C_a)_{ij}=\delta_{ia}\delta_{aj}-\delta_{i,a+1}\delta_{a+1,j}\,,
\ee
where $C_a$ are Cartan generators for $SU(N)$. It is easy to verify that the above Ansatz automatically satisfies the complex equations \eqref{HF}. Since the gauge field strength takes the form $F=-\partial\bar{\partial}f_aC_a$, the real equations \eqref{HD} translate into a Toda-like system of second-order partial differential equations on the $w$-plane for the $N-1$ functions $f_a$ \cite{Cecotti:2010bp}:
\be\label{Toda}
\partial\bar{\partial}f_a=a(N-a)e^{C_{ab}f_b}\,.
\ee

This vacuum configuration does not describe a system of intersecting D7-branes, because $\Phi$ cannot be diagonalized. It is, instead, the easiest instance of a T-brane, where the worldvolume flux and the non-commuting worldvolume scalars cannot be disentangled and permeate the entire brane worldvolume. Indeed, as explained in \cite{Cecotti:2010bp}, the solution described above originates from the following \emph{constant} nilpotent Higgs field in the so-called ``holomorphic'' gauge\footnote{In this gauge, all non-holomorphic data, like the $(1,1)$ component of the field strength, are invisible, and moreover the gauge connection can be gauged away. Supersymmetric vacua are thus simply specified by a holomorphic Higgs field modulo complexified gauge transformations.}
\be\label{Phih}
\Phi^{\rm h}=\left(\begin{array}{cccccc} 0&\phi_1^{\rm h}& & &\\ &0 & \phi_2^{\rm h} & & 0 \\ & & \ddots&\ddots& \\ &0&&0&\phi^{\rm h}_{N-1}\\&&&&0 \end{array}\right)\,,\qquad \phi^{\rm h}_a=\sqrt{a(N-a)}\,,
\ee
where the superscript $^{\rm h}$ stands for holomorphic. In the language of nilpotent orbits, this solution corresponds to the principal orbit of the algebra $sl(N,\C)$, and hence completely breaks the gauge group. Although we will only be explicitly discussing this solution throughout the paper, the generalization to the other nilpotent orbits of $sl(N,\C)$ is straightforward. The field $\Phi^{\rm h}$ will in general have a Jordan block structure, which reflects the partition of $N$ associated to the given nilpotent orbit. We can then repeat for each block a story analogous to the one we are about to tell for the single-block solution \eqref{Phih}, by simply replacing $N$ by the size of the block.

To conclude this section, let us remark that, in our T-brane tale, the real BPS equations \eqref{HD} are the ones that play the key role. Therefore, the physics we discuss cannot be captured by working in the holomorphic gauge.

\section{The D6 picture}\label{sec:D6}

In this section we would like to tell the first episode of our T-brane tale, which consists in taking a T-duality in the $w$-plane of the configuration introduced in the previous section. We will obtain a type IIA configuration of $N$ coincident D6-branes with non-commuting worldvolume scalars. To do that, we must first compactify one of the real directions of the $w$-plane, and require all vacuum profiles to be independent of it. Therefore let $w=\sigma+is$ be a cylindrical coordinate, with $s\simeq s+2\pi$ the compact direction. Having an isometry along $s$ amounts to reducing the Toda-like system \eqref{Toda} to an easier set of second-order ordinary differential equations:
\be\label{Toda1d}
f_a^{\prime\prime}(\sigma)=4a(N-a)e^{C_{ab}f_b(\sigma)}\,,
\ee
where the symbol $^\prime$ denotes the derivative with respect to $\sigma$. Note that, if we write
\be
f_a(\sigma) \equiv a(N-a) \log g(\sigma)\,,
\ee
the $N-1$ equations \eqref{Toda1d} collapse to a single one for $g(\sigma)$:
\be\label{MasterEq}
\frac{\rm d^2}{{\rm d}\sigma^2}\log g(\sigma)=4 g^2(\sigma)\,.
\ee
The gauge field \eqref{GaugeField} is sent by T-duality to a D6 worldvolume scalar, which we call $\Phi_3$ and whose eigenvalues parametrize the position of the various components of the D6 stack along the direction dual to $s$. Denoting $\Phi_1=\Phi+\Phi^\dagger$ and $\Phi_2=i(\Phi^\dagger-\Phi)$, we consider the following Ansatz
\be\label{WVScalarsD6}
\Phi_1(\sigma)=g(\sigma)\Sigma_1\,,\qquad\Phi_2(\sigma)=g(\sigma)\Sigma_2\,,\qquad\Phi_3(\sigma)=-\frac{g^\prime(\sigma)}{2g(\sigma)}\Sigma_3\,,
\ee
where $\Sigma_i$ are generalized Pauli matrices, defined as $\Sigma_1=\Phi^{\rm h}+\Phi^{{\rm h}\dagger}$, $\Sigma_2=i(\Phi^{{\rm h}\dagger}-\Phi^{\rm h})$ and $\Sigma_3={\rm diag}(N+1-2m)_{m=1,\ldots,N}$. The latter determine a $N\times N$ representation of the $su(2)$ algebra, as they satisfy the relation $[\Sigma_i,\Sigma_j]=2i\epsilon_{ijk}\Sigma_k$. It is immediate to see that, thanks to \eqref{MasterEq}, the worldvolume scalars \eqref{WVScalarsD6} satisfy the so-called Nahm equations \cite{Nahm:1979yw}:
\be\label{Nahm}
\frac{\rm d}{{\rm d}\sigma}\Phi_i(\sigma)=\frac{i}{2}\epsilon_{ijk}[\Phi_j(\sigma),\Phi_k(\sigma)]\,,
\ee
which are indeed the T-dual version of the Hitchin equations \eqref{HD}.

As explained in \cite{Constable:1999ac}, non-constant/non-commutative solutions of \eqref{Nahm}, just like those of \eqref{WVScalarsD6}, represent $1/4$ BPS vacuum configurations of a stack of D6-branes. Indeed, given that this solution is static, one can immediately derive its energy from the full non-Abelian DBI action of a stack of $N$ D6-branes extended along $\sigma$. Remarkably, this energy density can be expressed as the square root of a sum of perfect squares \cite{Constable:2001kv}:
\bea\label{DBID6}
E&\propto&\int{\rm d}\sigma\, {\rm STr}\sqrt{(\Phi_i^\prime-\tfrac{i}{2}\epsilon_{ijk}[\Phi_j,\Phi_k])^2+(1+\tfrac{i}{2}\epsilon_{ijk}\Phi_i^\prime[\Phi_j,\Phi_k])^2}\nonumber\\
&\geq&\int{\rm d}\sigma\, {\rm STr}\,(1+\tfrac{i}{2}\epsilon_{ijk}\Phi_i^\prime[\Phi_j,\Phi_k])\,,
\eea
where ${\rm STr}$ denotes the symmetrized trace \cite{Tseytlin:1997csa},\!\!\cite{Taylor:1999pr},\!\!\cite{Myers:1999ps}. Thus any solution of \eqref{MasterEq}-\eqref{WVScalarsD6} sets the first square to zero and, as one expects for supersymmetric solutions, also satisfies the full non-Abelian equations of motion \cite{Constable:1999ac}.

Let us examine the profile of the configuration given by \eqref{WVScalarsD6}. Note that, if we multiply both sides of equation \eqref{MasterEq} by $2(\log g)^{\prime}$, we can easily realize that the combination $\sqrt{(g^\prime/g)^2-4g^2}$ must be a constant, which we call $C$ and whose physical meaning will soon be apparent. Equation \eqref{MasterEq} thus reduces to the first-order equation
\be
g^\prime(\sigma)=- g(\sigma)\sqrt{4 g^2(\sigma)+C^2}\,.
\ee
Fixing the integration constant in such a way that the domain of definition of the solution is $\sigma>0$, we get
\be\label{Sol:g}
g(\sigma)=\frac{C}{2\sinh(C\sigma)}\,.
\ee

This scalar configuration describes a fuzzy funnel, in which the $N$ D6-branes, which are very close to each other at large $\sigma$, start expanding and eventually open up into a D8-brane located at $\sigma=0$. The cross-section of this funnel
at fixed $\sigma$ is a prolate ellipsoid of revolution (a rugby ball), which has two equal semi-axes ($\rho$) and a third longer semi-axis ($r$), given by
\be\label{eq:D6radii}
\begin{split}
\rho(\sigma)&=\sqrt{\frac{3{\rm Tr}(\Phi_1^2+\Phi_2^2)}{2N}}=\frac{C\,\sqrt{N^2-1}}{2\sinh(C\sigma)}\,,\\
r(\sigma)&=\sqrt{\frac{3{\rm Tr}\,\Phi_3^2}{N}}=\frac{C\,\sqrt{N^2-1}}{2\tanh(C\sigma)}\,.
\end{split}
\ee
The rugby-ball ellipsoid becomes more and more like a basketball as we approach the D8-brane, and the radius of this basketball blows up like $1/\sigma$. On the contrary, near the core of the funnel (for large $\sigma$), the small semi-axes, $\rho$, vanish exponentially, whereas the long semi-axis, $r$, goes to the constant $C\sqrt{N^2-1}/2$. This clarifies the meaning of $C$: Indeed, looking at the profile of $\Phi_3$ in \eqref{WVScalarsD6} for $\sigma\to+\infty$, we see that the $N$ D6-branes are uniformly distributed along the direction $3$ (the T-dual of $s$), with spacing given asymptotically by $C$. In the T-dual type IIB configuration of the previous section, a non-zero $C$ corresponds to a non-trivial asymptotic value of the Wilson line along $s$ for the stack of D7-branes. In the limit $C\to0$, our ellipsoid solution becomes the sphere solution of \cite{Constable:1999ac}, with radius $\rho=r=\sqrt{N^2-1}/(2\sigma)$, which collapses at the core of the funnel.

Finally, we can also calculate the energy density of this D6 configuration, by plugging the solution \eqref{WVScalarsD6} and \eqref{Sol:g} into (\ref{DBID6}). We find:
\begin{equation}\label{EnDensD6}
  \mathcal{E}(\sigma) \propto N + \frac{N(N^2 - 1)C^4}{4 \sinh^4 (C \sigma)} \left( 1 + \frac{2}{3} \sinh^2(C \sigma)\right)\,.
\end{equation}

For large $\sigma$, the non-commutative picture just outlined is the accurate description of the physical system at hand, because the D6-branes are all very close to each other and the physics is non-Abelian. However, if $C$ is larger than the string scale, the W-bosons are already massive at $\sigma=\infty$, the gauge group is broken to the Cartan, and no gluing mode can condense to give rise to a T-brane. This is confirmed by the exponential fall-off of the nilpotent profile of $\Phi$ for large $C$. Nevertheless, for very small values of $C$, the off-diagonal degrees of freedom are approximately massless and can acquire a non-trivial vacuum expectation value. We will encounter again the smallness condition for the constant $C$ when discussing validity regimes more systematically in Section \ref{sec:conc}.

\section{The D8 picture}\label{sec:D8}

When the number of D6-branes and the vevs of the non-Abelian fields are very large, the physics of the brane configuration considered in the previous section is captured by a different system, consisting of a single funnel-shaped D8-brane with non-trivial worldvolume gauge flux, $F_2$ \cite{Constable:1999ac}.\footnote{The Abelian and the non-Abelian brane pictures are equivalent in the large $N$ limit (up to $1/N^2$ corrections), and their regimes of validity are somewhat complementary, with an overlap whose size grows with $N$ (for more details see Section \ref{sec:conc}).} Hence, this can be thought of as a type of Myers effect for the D6-branes, with the difference that what keeps the D6-branes polarized are the boundary conditions, and not the bulk fluxes as in \cite{Myers:1999ps}. When all the non-commutative matrices in \eqref{WVScalarsD6} are equal, the corresponding Abelian D8-brane configuration is the Callan-Maldacena solution \cite{Callan:1997kz}, in which the boundary of the D6-branes ending on the D8-brane is viewed as a magnetic source for the gauge field living on the D8. In this section we will present the analogous construction for the more general ellipsoidal solution \eqref{eq:D6radii}, and this is the next episode of our T-brane tale.

Given the absence of any Ramond-Ramond background fields, the physics of our D8-brane is entirely determined by its DBI action
\begin{equation}\label{DBID8}
 S_{\textrm{DBI}} \propto \int \d^6 x \int \d\theta\d\varphi\d\sigma \sqrt{\det\left\{\delta|_{\rm D8} + F_2 \right\}} \,,
\end{equation}
where $\delta|_{\rm D8}$ denotes the pull-back of the flat metric on the D8 worldvolume and the $\d^6 x$ is the $\mathbb{R}^{5,1}$ measure. We will choose the following embedding of the D8 into space-time, tailored to the rugby-ball ellipsoidal shape (\ref{eq:D6radii}):
\begin{equation}\label{eq:embed}
  \begin{split}
    X^1 &= \rho(\sigma) \sin \theta \cos \varphi\,,\\
    X^2 &= \rho(\sigma) \sin \theta \sin \varphi\,,\\
    X^3 &= r(\sigma) \cos \theta\,,\\
    X^4 &= \sigma\,,
  \end{split}
\end{equation}
where $X^i$ are space-time coordinates, and $ \theta\,,\varphi\,,\sigma$ are worldvolume coordinates. The non-commutative coordinates $\Phi^1, \Phi^2, \Phi^3$ of the D6 picture have as commutative analogs $X^1, X^2, X^3$ respectively. We also restrict to a worldvolume flux corresponding to uniformly distributed D6 charge
\begin{equation}\label{eq:F2}
  F_2 = \frac{N}{2}\sin \theta\, \d \varphi \w \d \theta\,.
\end{equation}
With \eqref{eq:embed} and \eqref{eq:F2}, we can compute the $3\times3$ matrix appearing in \eqref{DBID8}:
\be\label{G+FSpheroid}
\delta|_{\rm D8} + F_2 = \left(\begin{array}{ccc} \rho^2 \cos^2\theta+r^2\sin^2\theta&-\tfrac{N}{2}\sin\theta &\sin\theta\cos\theta(\rho\rho^\prime-rr^\prime)\\ \\ \tfrac{N}{2}\sin\theta&\rho^2 \sin^2\theta &0 \\ \\ \sin\theta\cos\theta(\rho\rho^\prime-rr^\prime)&0&1+\rho^{\prime\, 2}\sin^2\theta+r^{\prime\, 2}\cos^2\theta\end{array}\right)\,.\vspace{.2cm}
\ee

\begin{figure}[h!]
  \begin{center}
    \includegraphics[scale=0.75]{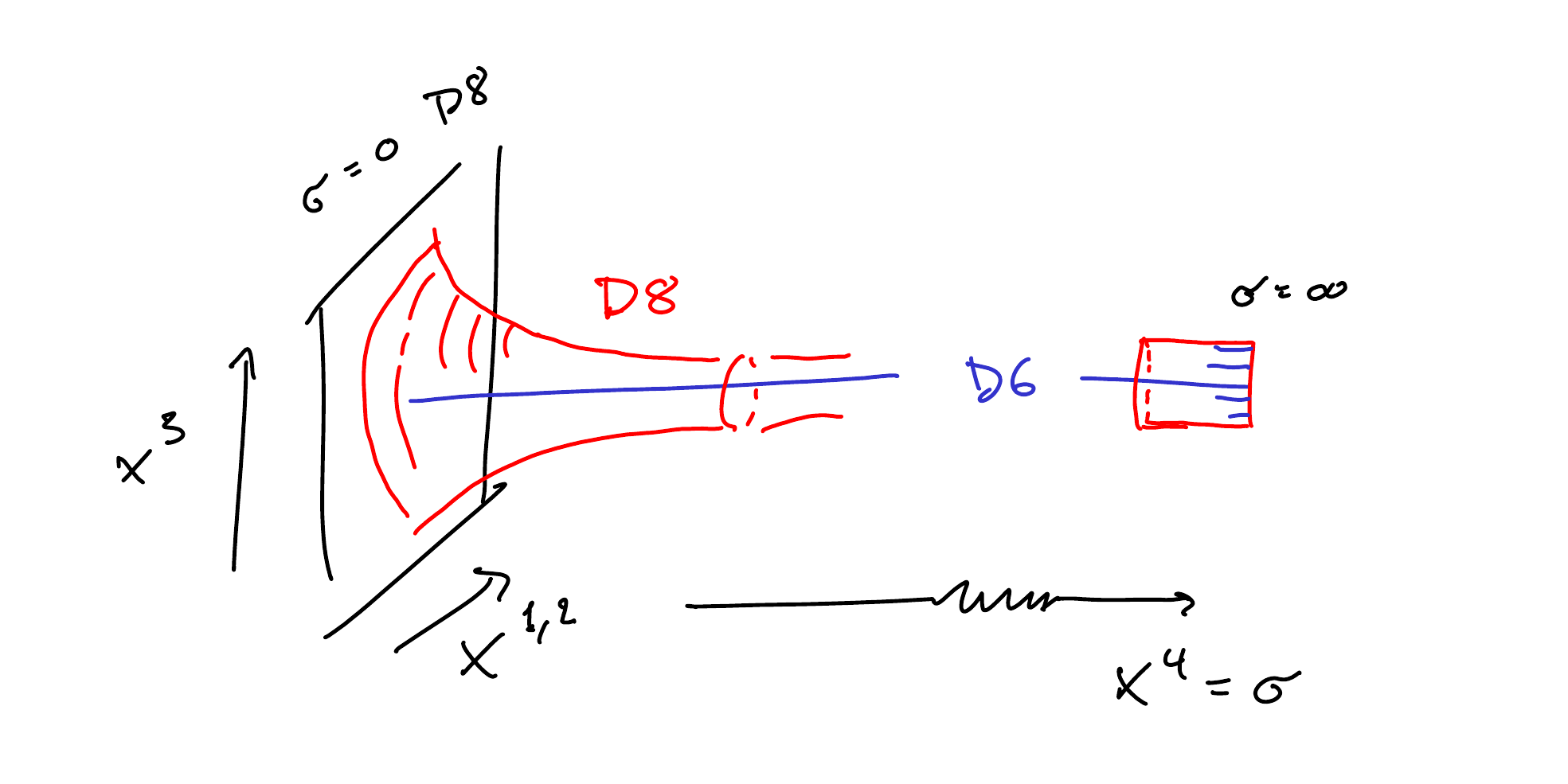}
  \end{center}
  \caption{Sketch of the D8 shape. The length of the line at $\sigma = \infty$, along $X^3$, is $CN$.\label{fig:D8shape}}
\end{figure}

The determinant of \eqref{G+FSpheroid} splits into a sum of three perfect squares, and setting two of them to zero (as we did for \eqref{DBID6}) gives the minimum-energy conditions for this system:
\be
\begin{split}
\rho^\prime&=-\frac{2}{N}\,\rho\, r\,,\\
r^\prime&=-\frac{2}{N}\,\rho^2\,.
\end{split}
\ee
These first-order differential equations are solved by:
\begin{equation}\label{eq:D8radii}
  \begin{split}
   \rho(\sigma) &= \frac{CN}{2\sinh(C\sigma)}\,,\\
    r(\sigma) &= \frac{CN}{2\tanh(C\sigma)}\,.
  \end{split}
\end{equation}
The shape of this D8-brane is sketched in Figure \ref{fig:D8shape}.

As expected, the expressions \eqref{eq:D8radii} agree with those in eq. \eqref{eq:D6radii}, up to $1/N^2$ terms\footnote{More precisely $\Delta\rho/\rho\sim\mathcal{O}(1/N^2)$, and the same for $r$.}.
Likewise, one can compute the energy density of the system and find agreement with the D6 picture, eq. \eqref{EnDensD6}, to the same level of approximation:
\be
\mathcal{E}(\sigma)\ \propto \ N+\frac{C^4N^3}{4 \sinh^4(C\sigma)}\left(1+\frac{2}{3}\sinh^2(C\sigma)\right)\,,
\ee
where we recognize the D6 contribution in the term linear in $N$ and the D8 contribution in the term cubic in $N$. In fact, the $C$-dependence of the second piece is illusory, as it disappears when integrating over $\sigma$ with a very small cutoff $\epsilon$ to avoid the divergence:
\be
E|_{\rm D8}\ \sim\ \left(\frac{N}{\epsilon}\right)^3\,.
\ee
In contrast to the D6 picture, the D8 picture is supposed to accurately describe the physics of this system in the region of space where both semi-axes are very large, and thus the curvature of the D8-brane is low in string units (see the discussion in Section \ref{sec:conc}).

Let us close this section by noting that the Ansatz \eqref{eq:embed} for the D8 embedding can be generalized further to a tri-axial ellipsoidal shape. To the best of our knowledge, all these supersymmetric solutions with ellipsoidal symmetry are new in the literature and interesting in their own right. More details about them, including their BPS equations and solutions thereof, are given in Appendix \ref{app:D8}, which thus contains the derivation of \eqref{eq:D8radii} as a particular limit.

\section{\ldots~and back again}\label{sec:back}

Having derived the D8-brane shape in the previous section, we are approaching the epilogue of our T-brane tale: We return to type IIB by T-dualizing the Abelian D8-brane configuration back along the $X^3$ direction. The result will be a single D7-brane with vanishing worldvolume flux, extended along a specific holomorphic curve in $\mathbb{C}^2$.

As emphasized in Section \ref{sec:D8}, the D8 picture is reliable when the semi-axes \eqref{eq:D8radii} are very large in string units. Expanding the rugby-ball solution in this region of space (small $\sigma$), we obtain $\rho \approx r \approx {N}/{2\sigma}$.

In order to see the fate of our funnel-shaped D8-brane after T-duality along $X^3$, the first step is to change coordinates on the D8 worldvolume, trading $\theta$ for $X^3$. Remebering that $\rho \approx r$, the matrix \eqref{G+FSpheroid} becomes
\be\label{DBImatrixX3}
\delta|_{\rm D8} + F_2 \approx \left(\begin{array}{ccc}\frac{1}{1-(X^3/\rho)^2}&\frac{N}{2\rho}& -\rho^\prime\frac{X^3/\rho}{1-(X^3/\rho)^2}\\ \\ -\frac{N}{2\rho} & \rho^2(1-(\frac{X^3}{\rho})^2) & \frac{N\rho^\prime}{2\rho}\frac{X^3}{\rho} \\ \\  -\rho^\prime\frac{X^3/\rho}{1-(X^3/\rho)^2}& -\frac{N\rho^\prime}{2\rho}\frac{X^3}{\rho}  & 1+\frac{\rho^{\prime\,2}}{1-(X^3/\rho)^2} \end{array}\right)\,.\vspace{.2cm}
\ee
Even though this D8 configuration displays no isometry along $X^3$, it is not hard to see that the determinant of the above matrix, and thus the DBI dynamics of the system, does not depend on $X^3$. In order to be able to T-dualize this solution, we zoom in on the equator of the rugby ball, $\theta = \pi/2$, which eliminates the $X^3$ dependence of \eqref{DBImatrixX3}, and thus is equivalent to approximating the rugby ball with a cylinder.
The profile $\rho=\rho(\sigma)$ is unaffected by this operation, because the equation of motion remains the same.
Very roughly, this happens because, at each fixed $\sigma$, the local D8 charge contributions on the $X^{1,2}$-plane above the equator cancel out the corresponding ones below the equator.

One may worry that zooming in on the equator discards information regarding the ends of the rugby ball. In fact, one could have raised a similar objection at the first step of the tale, where we T-dualized the D7 stack but considered D6-branes spread over a finite interval of the T-duality direction.\footnote{Since we T-dualized at the level of the equations of motion, we did not discuss this issue at that stage.} However, in the D6 picture, the type IIB T-brane data should be captured just by the D6-branes in the middle of the interval (where there is an approximate isometry along the T-duality direction), and not by the D6-branes near the boundaries. Zooming in on the equator implements this in the Abelian D8 picture.

\begin{figure}[h!]
  \begin{center}
    \includegraphics[scale=0.75]{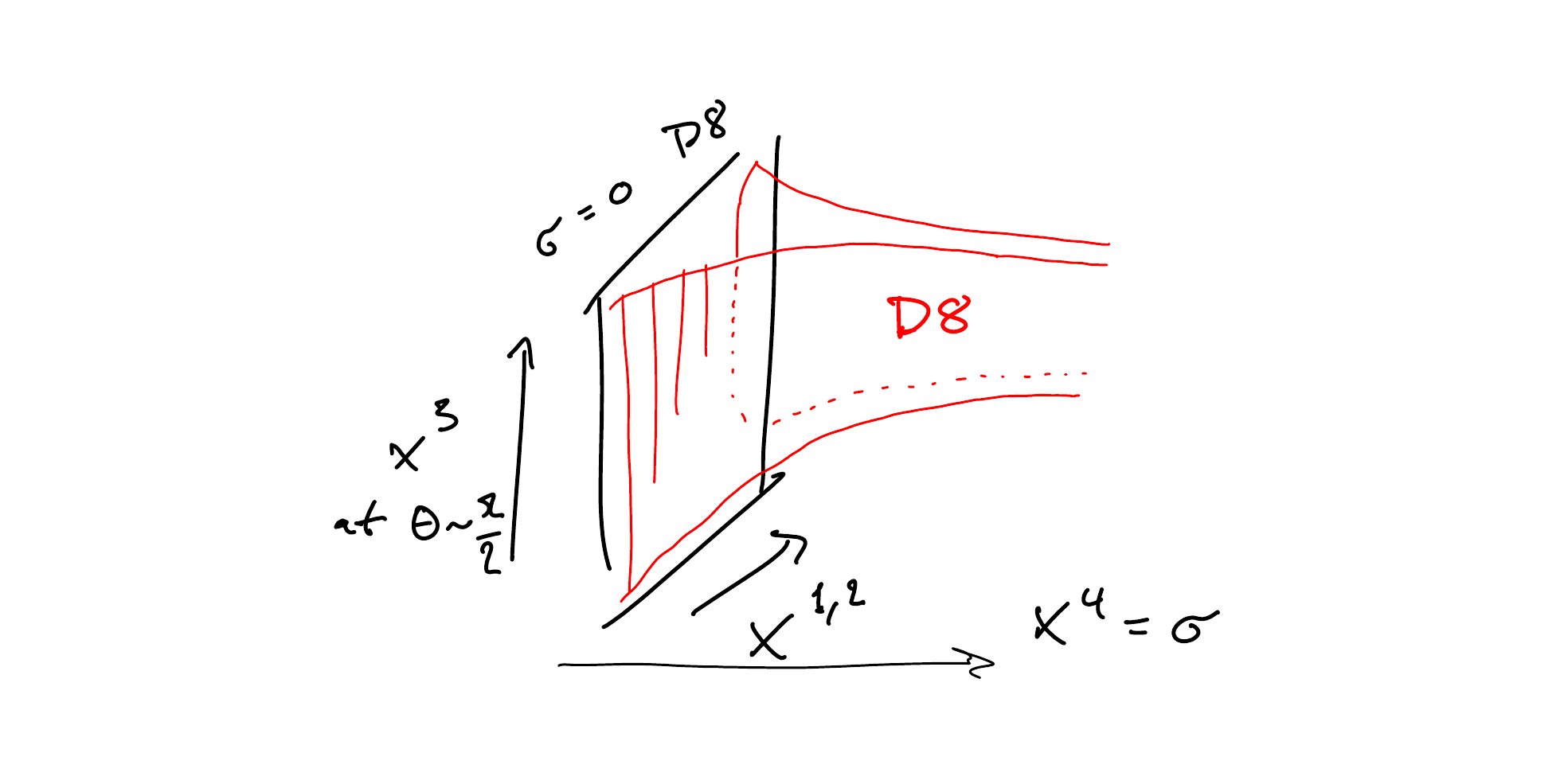}
  \end{center}
  \caption{Sketch of the D8 shape after zooming into a small neighborhood of the equator.\label{fig:D8shapezoom}}
\end{figure}

Therefore we are led to consider a D8-brane whose shape is sketched in Figure \ref{fig:D8shapezoom} and whose defining data are the following
\be\begin{split}
\rho&=\frac{N}{2\sigma}\,,\\
F_{\varphi\,3}&= \partial_{\varphi} A_3= -\frac{N}{2\rho}=-\sigma\,.
\end{split}\ee
T-dualizing the above configuration along $X^3$ is straightforward: We obtain a D7-brane with no worldvolume flux and with shape determined by the following pair of equations in the four-dimensional ambient space parametrized by $\rho,\varphi,\sigma,\tilde{X}^3$:
\be\begin{split}\label{RealD7}
\rho&=\frac{N}{2\sigma}\,,\\
\varphi&=-\frac{\tilde{X}^3}{\sigma}\,.
\end{split}\ee
where $\tilde{X}^3$, which the potential $A_3$ is mapped to, denotes the coordinate T-dual to $X^3$.

\begin{figure}[h!]
  \begin{center}
    \includegraphics[scale=0.5,trim={50 50 35 50},clip]{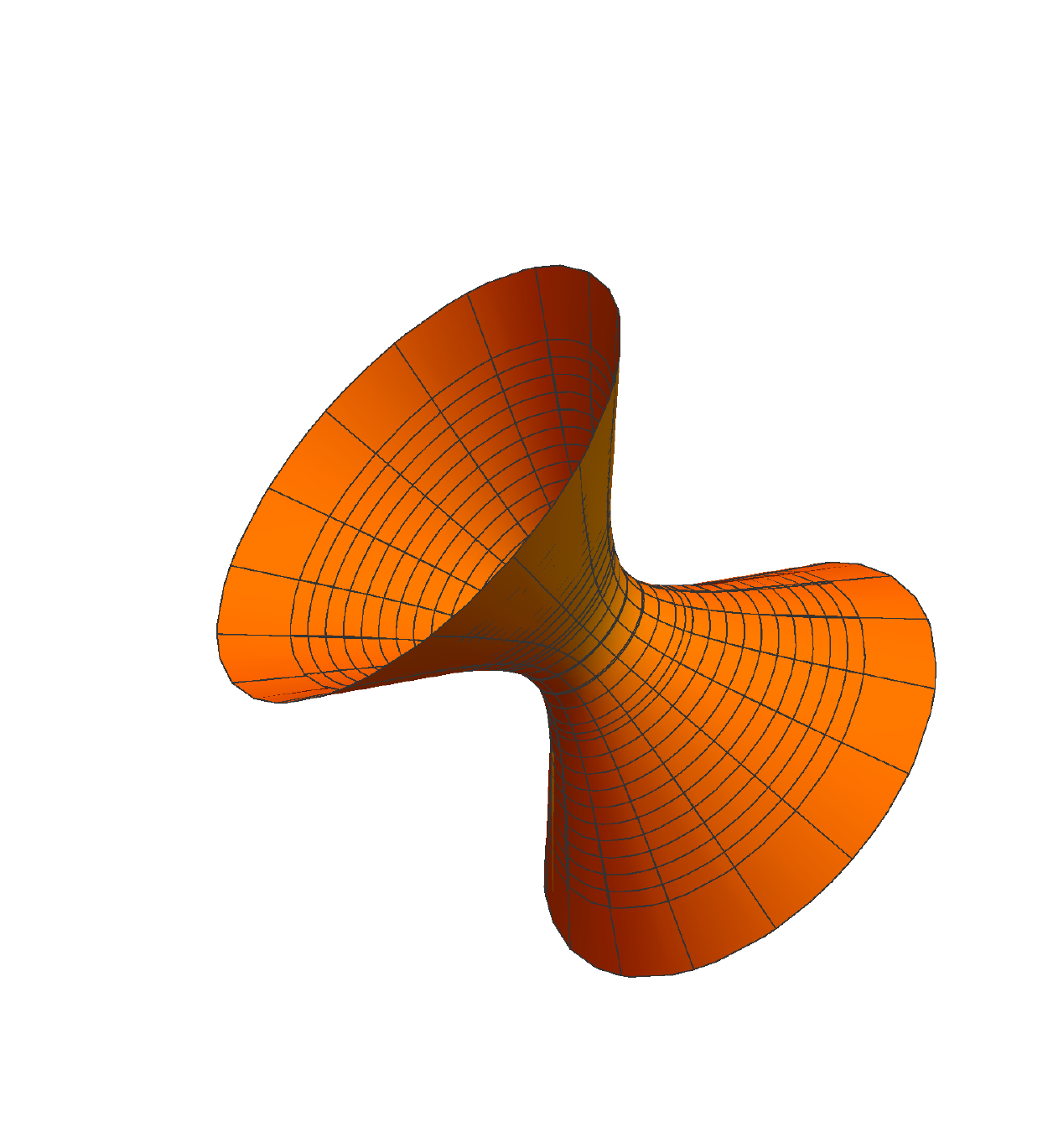}
  \end{center}
  \caption{Three-dimensional profile of the D7-brane after T-duality, in which $|Z|$ and $|W|$ are plotted and rotated on the diagonal. The size of the minimal circle is proportional to $N$.\label{fig:Z1Z2}}
\end{figure}

In order to make supersymmetry manifest, we can translate the two real equations \eqref{RealD7} into a single complex holomorphic one. Consider an ambient $\mathbb{C}^2$ parametrized by the complex coordinates
\begin{equation}\label{ComplexCoord}
  \begin{split}
    Z &= \rho \,e^{i \varphi}\,,\\
    W &= \sigma\, e^{i \tilde{X}^3/\sigma}\,.
  \end{split}
\end{equation}
Note that the coordinate $Z$ is identical to the $z$ used in Section \ref{sec:T}, whereas $W$ differs from the $w$ of Section \ref{sec:T}. However, for $\sigma\gg1$, $W\approx \sigma+i\tilde{X}^3=w$, where we identified $\tilde{X}^3\equiv s$.
Using \eqref{RealD7} and \eqref{ComplexCoord}, our D7-brane can then be seen to wrap the holomorphic curve
\begin{equation}\label{FinalEq}
  Z W = \frac{N}{2}\,.
\end{equation}
A three-dimensional plot of this shape is displayed in Figure \ref{fig:Z1Z2}.

We have therefore reached an alternative description of our starting T-brane configuration \eqref{Phiu}, whereby the flux data of the latter is now encoded in the curvature of the worldvolume surface \eqref{FinalEq}, whose Ricci scalar is given by
\begin{equation}\label{RicciSc}
    R = -256 N^2 \frac{\sigma^6}{(N^2 + 4 \sigma^4)^3}\,.
\end{equation}
Remarkably, even if our derivation of \eqref{FinalEq} was performed in the regime of small $\sigma$, where the Ricci curvature of the D8-brane is small, our final D7-brane configuration never suffers from a large-curvature problem. Indeed, as manifest from \eqref{RicciSc}, working at large $N$ is enough to guarantee that our D7-brane has a small curvature for all values of $\sigma$. Thus $\alpha'$ corrections are negligible, which suggests that our Abelian picture of the T-brane could give a correct description of the physics even for small vevs of the non-Abelian fields. We plan to investigate this exciting possibility in the future.

\section{Discussion}\label{sec:conc}

In this paper we have considered a specific class of T-brane solutions with $5+1$-dimensional Poincar\'e invariance, realized as a particular set of eight-supercharge  vacua on a stack of D7-branes with non-commuting worldvolume scalars and non-trivial worldvolume flux. We have worked in a local patch of the internal part of the D7-brane worldvolume and of its transverse space, thus decoupling gravitational effects and neglecting all issues related to compactification.

The class of T-branes we have focused on is specified by an integer partition of $N$, the number of D7-branes composing the stack. When $N$ is large, we have found a novel description of the T-brane corresponding to the maximal partition, consisting of a single D7-brane with no worldvolume flux, wrapping the smooth holomorphic curve $ZW=N/2$  in $\C^2$.
One can trivially extend our construction to a generic partition of $N$, $\{n_i\}_{i=1,\ldots,k}$. When all the $n_i$ are large, our new description will consist of $k$ fluxless D7-branes wrapping the curves $\{ZW=n_i/2\}_{i=1,\ldots,k}$. In particular, we observe an enhancement of the preserved gauge group when two or more $n_i$ coincide, which is consistent with the original T-brane description. The maximal partition that we have analyzed corresponds to a vacuum which breaks the $SU(N)$ gauge symmetry completely.

We have achieved our result by making a detour through type IIA string theory, and relating the T-branes to certain supersymmetric systems of D$p$-branes ending on D$(p+2)$-branes that give rise to rugby-ball-shaped bion cores.

Throughout our T-brane tale, we have disregarded the backreaction of all D-branes on the bulk geometry, by tuning the string coupling constant $g_s$ such that $g_sN\ll1$. This weak coupling limit has no effect on the funnel shapes we have discussed, because they are determined by the interplay between the D6 and D8 tensions, and hence are independent of $g_s$ \cite{Callan:1997kz}.

We have argued that our Abelian description of the T-brane is the correct one when vevs of the non-Abelian fields and flux densities are large in string units, and hence the non-Abelian description \cite{Cecotti:2010bp}, based on equation \eqref{HD}, breaks down.

Indeed, in the T-dual type IIA description, the Abelian and the non-Abelian pictures have complementary regimes of validity. However, as pointed out in \cite{Constable:1999ac}, these two regimes are expected to overlap, in a region whose size grows as $N$ increases. To see this for our rugby-ball D6-D8 funnels, we have to analyze when higher-order variations of the worldvolume fields are much smaller than first-order variations, which are the only ones captured by the DBI action. For the non-Abelian physics of the D6-branes, this condition is
\be
|\Phi_i^{\prime\prime}|\ll|\Phi_i^\prime|\,,\qquad \forall\, i=1,2,3\,.
\ee
Using the explicit solution \eqref{WVScalarsD6} and \eqref{Sol:g}, it is easy to express this as an upper bound for the larger semi-axis: $r\ll N$. Since, at $\sigma=\infty$, $r$ approaches a non-zero minimal value, $r_{\rm min}\sim CN$, the non-Abelian description is consistent when $C\ll1$. We thus recover a condition that we guessed on physical grounds at the end of Section \ref{sec:D6}.
The Abelian D8 picture is valid when (in string units)
\be\label{ValidityD8}
\left|\frac{{\rm d^2}\sigma}{{\rm d}\rho^2}\right|\ll\left|\frac{{\rm d}\sigma}{{\rm d}\rho}\right|\,,\qquad
\left|\frac{{\rm d^2}\sigma}{{\rm d}r^2}\right|\ll\left|\frac{{\rm d}\sigma}{{\rm d}r}\right|\,.
\ee
By inspecting the solution \eqref{eq:D8radii}, we easily see that both these conditions are satisfied so long as $\rho\gg CN$, which gives a lower bound for the smaller semi-axis. Therefore the quantity $CN$ should not be too large, in order not to spoil the D8 picture. Thus, when $C$ is of order $1/N$, both the Abelian and the non-Abelian descriptions are valid, and the overlap is proportional to $N$.\footnote{This is similar to the analysis made in \cite{Constable:1999ac} for the spherical solution.}

Based on this analysis, one would expect that our Abelian description of the T-brane is not only valid for very large non-Abelian vevs and flux densities (when the non-Abelian description breaks down), but also when these vevs are small and the non-Abelian description is reliable. In fact, the regime of validity of the Abelian T-brane description appears to extend over the whole non-Abelian regime of validity: This is because for large $N$ the curvature \eqref{RicciSc} is low for the whole range of $\sigma$ and there is no worldvolume flux. We believe that this remarkable phenomenon deserves a deeper understanding.\footnote{For instance, equation \eqref{HD} is know to receive $\alpha'$ corrections \cite{Minasian:2001na}, which our Abelian picture should automatically encode.}

As mentioned in the Introduction, our Abelian description of T-branes is reminiscent of brane recombination for T-branes associated to ``reconstructible'' Higgs fields\footnote{A Higgs field is called reconstructible if its spectral surface is non-singular \cite{Cecotti:2010bp}.} \cite{Cecotti:2010bp}. Our Higgs field \eqref{Phiu} is non-reconstructible, and if one tries to carry over the brane recombination analysis of \cite{Cecotti:2010bp}, one finds a highly singular configuration \cite{Collinucci:2014qfa}. The effect that we find in this paper is a large-$N$ effect, which cannot be captured in a gauge where one keeps only the holomorphic data, as done in \cite{Cecotti:2010bp,Collinucci:2014qfa}. It would be interesting to understand if there is any connection between the effect we find and brane recombination, and whether our solution can be thought of as the smoothing out of the singular ``brane-recombination'' shape of \cite{Collinucci:2014qfa} by non-holomorphic physics.

The main result of this paper raises the obvious question of whether there is a direct way of deriving our Abelian description of T-branes without the need of following an indirect path, like in Figure \ref{MasterFigure}. A positive answer would open up a plethora of new research directions. For example, it would allow one to explore more complicated T-brane solutions, such as those involving non-constant holomorphic functions in the (holomorphic) Higgs field \eqref{Phih}. It would also be very exciting, especially in light of possible phenomenological applications, to extend our derivation to four-dimensional T-brane vacua with four supercharges, possibly containing non-trivial monodromies.

\section*{Acknowledgements}

We would like to thank I\~naki Garc\'ia-Etxebarria, Alessandro Tomasiello and Gianluca Zoccarato for helpful discussions. 
The work of I.B. and J.B. was supported by the John Templeton Foundation Grant 48222. The work of J.B. was also supported by the CEA Eurotalents program. The work of R.M. was supported by the ANR grant 12-BS05-003-01. The work of R.S. was supported by the ERC Starting Grant 259133 - ObservableString. R.M. and R.S. would like to thank the Institut Henri Poincar\'e for hospitality during this work.

\appendix

\section{Funnel shapes for the D8-branes}\label{app:D8}

In this appendix, we would like to describe a class of $1/4$ BPS solutions consisting of funnel-shaped D8-branes whose cross-section is a generic tri-axial ellipsoid. They naturally arise as solutions of the Abelian DBI dynamics of a single D8-brane with magnetic flux. To the best of our knowledge they are new in the literature, and reduce to the long-known spherical solution of \cite{Callan:1997kz} for a particular choice of parameters. The more general bi-axial ellipsoid solutions (rugby balls), relevant for this paper and presented in Section \ref{sec:D8}, can be recovered as a limit of these most general solutions.

Let us start with the following tri-axial Ansatz for the embedding of a D8-brane with a worldvolume parametrized by $\theta,\varphi,\sigma$, into a flat Euclidean ambient space with coordinates $X^1,\ldots,X^4$:
\begin{equation}
  \begin{split}
    X^1 &= r_1(\sigma) \sin \theta \cos \varphi\,,\\
    X^2 &= r_2(\sigma) \sin \theta \sin \varphi\,,\\
    X^3 &= r_3(\sigma) \cos \theta\,,\\
    X^4 &= \sigma\,.
  \end{split}
\end{equation}
We also take a worldvolume gauge field strength, $F_2 = \frac{N}{2} \sin \theta \d \varphi \w \d \theta$, that is uniform\footnote{Solutions with non-uniform $F_2$ can in principle also be constructed. See \cite{Bak:2006yb} for some examples involving D2-branes.}.  The worldvolume flux and the pull-back of the metric give the $3\times3$ matrix:\vspace{.2cm}
\begin{equation}
  \begin{split}
    &\delta|_{\textrm{D8}} + F_2 = \\ &{\tiny
  \begin{pmatrix}
   \cos ^2\theta \left(r_2^2 \sin ^2\varphi+r_1^2 \cos ^2\varphi \right)+r_3^2 \sin ^2\theta & -\frac{1}{2} \sin \theta
     \left(\text{N}+\left(r_1^2-r_2^2\right) \cos \theta \sin (2 \varphi
     )\right) & \sin \theta \cos \theta \left(r_2 r_2' \sin ^2(\varphi
     )+r_1 r_1' \cos ^2\varphi-r_3 r_3'\right) \\
   \frac{1}{2} \sin \theta \left(\text{N}+\left(r_2^2-r_1^2\right) \cos
     \theta \sin (2 \varphi )\right) & \sin ^2\theta \left(r_1^2 \sin
     ^2\varphi+r_2^2 \cos ^2\varphi\right) & \sin ^2\theta \left(r_2
     r_2'-r_1 r_1'\right) \sin \varphi \cos \varphi \\
   \sin \theta \cos \theta \left(r_2 r_2' \sin ^2\varphi+r_1 r_1' \cos
     ^2\varphi-r_3 r_3'\right) & \sin ^2\theta \left(r_2 r_2'-r_1
     r_1'\right) \sin \varphi \cos \varphi & \sin ^2\theta
     \left(r_2'{}^2 \sin ^2\varphi+r_1'{}^2 \cos
     ^2\varphi\right)+\cos ^2\theta r_3'{}^2+1 \nonumber\\
  \end{pmatrix}.
  }
  \end{split}
\end{equation}
This results in a DBI Lagrangian density, whose square can be written as a sum of four perfect squares:
\begin{equation}
  \begin{split}
    \mathcal{L}^2_{\rm DBI} &\propto \sin ^2\varphi  \left(
      \frac{N}{2} - r_1' r_2 r_3 \sin ^2\theta
   \cos ^2\varphi -r_1 r_2' r_3 \sin ^2\theta  \sin ^2\varphi -r_1 r_2 r_3' \cos ^2\theta
    \right)^2\\
    &\quad + (\sin^2\theta \cos \varphi )^2 \left(\frac{1}{2} N r_1'+r_2 r_3\right)^2 + (\sin
   ^2 \theta \sin \varphi )^2 \left(\frac{1}{2} N r_2'+r_1 r_3\right)^2\\
    &\quad + (\sin\theta \cos\theta)^2 \left(\frac{1}{2} N r_3'+r_1 r_2\right)^2\,,
  \end{split}
\end{equation}
which makes it easy to extract the BPS equations as minimum-energy conditions
\begin{equation}
  r_{i+1}' = - \frac{2}{N} r_{i+2} r_{i+3}\,,\qquad \textrm{with }\; i \equiv i+3\,.
\end{equation}
These equations can be disentangled to give the following system:
\begin{equation}\label{eq:distangle}
  \begin{split}
    &r_1^2 - r_2^2 - \left( \frac{N C_2}{2} \right)^2 = 0\,,\\
    &r_1^2 - r_3^2 - \left( \frac{N C_3}{2} \right)^2 = 0\,,\\
    &r_1' + \frac{2}{N} \sqrt{r_1^2 - \left( \frac{N C_2}{2} \right)^2}\sqrt{r_1^2 - \left( \frac{N C_3}{2}\right)^2} = 0\,,
  \end{split}
\end{equation}
where $C_2$ and $C_3$ are integration constants. The only differential equation left in \eqref{eq:distangle} has the solution
\begin{equation}
  r_1 = \frac{N}{2} C_2\, \textrm{sn}\left( C_3  \left(\sigma - C_1 \right)\left| \left(\frac{C_2}{C_3}\right)^2 \right.\right)\,,
\end{equation}
where ``$\textrm{sn}$'' is the Jacobi sine elliptic function and $C_1$ is the remaining integration constant  \cite{Gradshteyn}. This function has poles at
\begin{equation}
  2 m_1 K + (2 m_2 + 1) i K'\,,
\end{equation}
where $m_i$ are integers, $K=K(k)$ and $K' = K(1-k)$, where $K$ is the Elliptic K function and $k = (C_2/C_3)^2$. This is only one pole, because the function has a period given by $4 m_1 K + 2 m_2 i K'$, and it is odd under shifts of the argument by $2 m_1 K$. Hence, by defining $C_4 = -C_1 + i K'/C_3$, we have the following leading behavior at $\sigma \approx C_4$:
\begin{equation}\label{eq:smallsigma}
  r_1 = \frac{N}{2(\sigma - C_4)} + \mathcal{O}(\sigma - C_4)\,.
\end{equation}
This pole represents a D8-brane which, without loss of generality, we can place at $\sigma = 0$, by fixing $C_4 = 0$. The expression for $r_1$ can now be written as
\begin{equation}
  r_1 = \frac{N C_3}{2\, \textrm{sn} \left( C_3 \sigma \left| \left( \frac{C_2}{C_3} \right)^2 \right.\right)}\,.
\end{equation}
This expression is generally periodic, and describes periodically reoccurring D8-branes. However, there are two limits in which the period becomes infinite:
\begin{equation}
  \begin{split}
    \lim_{C_2 \to 0} r_1   &= \frac{N\tilde{C}_3}{2 \sinh(\tilde{C}_3 \sigma)}\,,\\
    \lim_{C_2 \to C_3} r_1 &= \frac{NC_3}{2 \tanh({C}_3 \sigma)}\,,
  \end{split}
\end{equation}
where $\tilde{C}_3^2 = -C_3^2$. The first limit requires $C_3$ to be purely imaginary, while in the second limit $C_3$ is real. From (\ref{eq:distangle}) we can see that the first limit leads to $r_1 = r_2 \leq r_3$, hence the tri-axial ellipsoid is reduced to the rugby ball discussed in Section \ref{sec:D8}. The other limit is equivalent, but with $r_1 \geq r_2 = r_3$.

The first limit is the one used in Section \ref{sec:D8}, with $C_3 = C$, $r_1 = r_2 = \rho$ and $r_3 = r$. Its small-$\sigma$ behavior is given in \eqref{eq:smallsigma}, while for large $\sigma$ we have
\begin{equation}
  \begin{split}
    r &\to \frac{N}{2} C\,,\\
    \rho &\to N C e^{-C \sigma}\,.
  \end{split}
\end{equation}
The limit $C \to 0$ reproduces the spherical shape of \cite{Callan:1997kz,Constable:1999ac}
\begin{equation}
  r = \rho = \frac{N}{2\sigma}\,.
\end{equation}


\bibliography{refs}

\providecommand{\href}[2]{#2}\begingroup\raggedright\begin{thebibliography}{10}

\bibitem{Cecotti:2010bp}
S.~Cecotti, C.~Cordova, J.~J. Heckman and C.~Vafa,  {\em {T-Branes and
  Monodromy}}, JHEP {\bf 07} (2011) 030
[\href{http://www.arXiv.org/abs/1010.5780}{{\tt 1010.5780}}].

\bibitem{Donagi:2003hh}
R.~Donagi, S.~Katz and E.~Sharpe,  {\em {Spectra of D-branes with higgs vevs}},
  Adv. Theor. Math. Phys. {\bf 8} (2004), no.~5, 813--859
[\href{http://www.arXiv.org/abs/hep-th/0309270}{{\tt hep-th/0309270}}].

\bibitem{Chiou:2011js}
C.-C. Chiou, A.~E. Faraggi, R.~Tatar and W.~Walters,  {\em {T-branes and Yukawa
  Couplings}}, JHEP {\bf 05} (2011) 023
[\href{http://www.arXiv.org/abs/1101.2455}{{\tt 1101.2455}}].

\bibitem{Donagi:2011jy}
R.~Donagi and M.~Wijnholt,  {\em {Gluing Branes, I}}, JHEP {\bf 05} (2013) 068
[\href{http://www.arXiv.org/abs/1104.2610}{{\tt 1104.2610}}].

\bibitem{Donagi:2011dv}
R.~Donagi and M.~Wijnholt,  {\em {Gluing Branes II: Flavour Physics and String
  Duality}}, JHEP {\bf 05} (2013) 092
[\href{http://www.arXiv.org/abs/1112.4854}{{\tt 1112.4854}}].

\bibitem{Marsano:2012bf}
J.~Marsano, N.~Saulina and S.~Sch{\"a}fer-Nameki,  {\em {Global Gluing and
  $G$-flux}}, JHEP {\bf 08} (2013) 001
[\href{http://www.arXiv.org/abs/1211.1097}{{\tt 1211.1097}}].

\bibitem{Anderson:2013rka}
L.~B. Anderson, J.~J. Heckman and S.~Katz,  {\em {T-Branes and Geometry}}, JHEP
  {\bf 05} (2014) 080
[\href{http://www.arXiv.org/abs/1310.1931}{{\tt 1310.1931}}].

\bibitem{DelZotto:2014hpa}
M.~Del~Zotto, J.~J. Heckman, A.~Tomasiello and C.~Vafa,  {\em {6d Conformal
  Matter}}, JHEP {\bf 02} (2015) 054
[\href{http://www.arXiv.org/abs/1407.6359}{{\tt 1407.6359}}].

\bibitem{Collinucci:2014qfa}
A.~Collinucci and R.~Savelli,  {\em {T-branes as branes within branes}}, JHEP
  {\bf 09} (2015) 161
[\href{http://www.arXiv.org/abs/1410.4178}{{\tt 1410.4178}}].

\bibitem{Collinucci:2014taa}
A.~Collinucci and R.~Savelli,  {\em {F-theory on singular spaces}}, JHEP {\bf
  09} (2015) 100
[\href{http://www.arXiv.org/abs/1410.4867}{{\tt 1410.4867}}].

\bibitem{Cicoli:2015ylx}
M.~Cicoli, F.~Quevedo and R.~Valandro,  {\em {De Sitter from T-branes}}, JHEP
  {\bf 03} (2016) 141
[\href{http://www.arXiv.org/abs/1512.04558}{{\tt 1512.04558}}].

\bibitem{Carta:2015eoh}
F.~Carta, F.~Marchesano and G.~Zoccarato,  {\em {Fitting fermion masses and
  mixings in F-theory GUTs}}, JHEP {\bf 03} (2016) 126
[\href{http://www.arXiv.org/abs/1512.04846}{{\tt 1512.04846}}].

\bibitem{Collinucci:2016hpz}
A.~Collinucci, S.~Giacomelli, R.~Savelli and R.~Valandro,  {\em {T-branes
  through 3d mirror symmetry}},
\href{http://www.arXiv.org/abs/1603.00062}{{\tt 1603.00062}}.

\bibitem{Myers:1999ps}
R.~C. Myers,  {\em {Dielectric branes}}, JHEP {\bf 12} (1999) 022
[\href{http://www.arXiv.org/abs/hep-th/9910053}{{\tt hep-th/9910053}}].

\bibitem{Constable:1999ac}
N.~R. Constable, R.~C. Myers and O.~Tafjord,  {\em {The Noncommutative bion
  core}}, Phys. Rev. {\bf D61} (2000) 106009
[\href{http://www.arXiv.org/abs/hep-th/9911136}{{\tt hep-th/9911136}}].

\bibitem{Callan:1997kz}
C.~G. Callan and J.~M. Maldacena,  {\em {Brane death and dynamics from the
  Born-Infeld action}}, Nucl. Phys. {\bf B513} (1998) 198--212
[\href{http://www.arXiv.org/abs/hep-th/9708147}{{\tt hep-th/9708147}}].

\bibitem{Beasley:2008dc}
C.~Beasley, J.~J. Heckman and C.~Vafa,  {\em {GUTs and Exceptional Branes in
  F-theory - I}}, JHEP {\bf 01} (2009) 058
[\href{http://www.arXiv.org/abs/0802.3391}{{\tt 0802.3391}}].

\bibitem{Hitchin:1986vp}
N.~J. Hitchin,  {\em {The Selfduality equations on a Riemann surface}}, Proc.
  Lond. Math. Soc. {\bf 55} (1987)
59--131.

\bibitem{Nahm:1979yw}
W.~Nahm,  {\em {A Simple Formalism for the BPS Monopole}}, Phys. Lett. {\bf
  B90} (1980)
413--414.

\bibitem{Constable:2001kv}
N.~R. Constable, R.~C. Myers and O.~Tafjord,  {\em {Fuzzy funnels: NonAbelian
  brane intersections}}, in {\em {Strings 2001: International Conference
  Mumbai, India, January 5-10, 2001}}.
\newblock 2001.
\newblock
\href{http://www.arXiv.org/abs/hep-th/0105035}{{\tt hep-th/0105035}}.
\newblock

\bibitem{Tseytlin:1997csa}
A.~A. Tseytlin,  {\em {On nonAbelian generalization of Born-Infeld action in
  string theory}}, Nucl. Phys. {\bf B501} (1997) 41--52
[\href{http://www.arXiv.org/abs/hep-th/9701125}{{\tt hep-th/9701125}}].

\bibitem{Taylor:1999pr}
W.~Taylor and M.~Van~Raamsdonk,  {\em {Multiple Dp-branes in weak background
  fields}}, Nucl. Phys. {\bf B573} (2000) 703--734
[\href{http://www.arXiv.org/abs/hep-th/9910052}{{\tt hep-th/9910052}}].

\bibitem{Minasian:2001na}
R.~Minasian and A.~Tomasiello,  {\em {Variations on stability}}, Nucl. Phys.
  {\bf B631} (2002) 43--65
[\href{http://www.arXiv.org/abs/hep-th/0104041}{{\tt hep-th/0104041}}].

\bibitem{Bak:2006yb}
D.~Bak, N.~Ohta and P.~K. Townsend,  {\em {The D2 Susy zoo}}, JHEP {\bf 03}
  (2007) 013
[\href{http://www.arXiv.org/abs/hep-th/0612101}{{\tt hep-th/0612101}}].

\bibitem{Gradshteyn}
Gradshteyn and Ryzhik, {\em Table of Integrals, Series, and Products}.
\newblock Alan Jeffrey, Editor, fifth~ed., 1994.

\end{thebibliography}\endgroup

\bibliographystyle{utphysmodb}

\end{document}